# Weak-to-strong transition of quantum measurement in a trapped-ion system

Yiming Pan[ ][1,5 ✉], Jie Zhang[ ][2,3,5], Eliahu Cohen[ ][4], Chun-wang Wu[ ][2,3 ✉], Ping-Xing Chen[2,3] and Nir Davidson[ ][1]

**Quantum measurement remains a puzzle through its stormy history from the birth of quantum mechanics to state-of-the-art quantum technologies. Two complementary measurement schemes have been widely investigated in a variety of quantum systems: von Neumann's projective 'strong' measurement and Aharonov's weak measurement. Here, we report the observation of a weak-to-strong measurement transition in a single trapped $^{40}Ca^+$ ion system. The transition is realized by tuning the interaction strength between the ion's internal electronic state and its vibrational motion, which play the roles of the measured system and the measuring pointer, respectively. By pre- and post-selecting the internal state, a pointer state composed of two of the ion's motional wavepackets is obtained, and its central-position shift, which corresponds to the measurement outcome, demonstrates the transition from the weak-value asymptotes to the expectation-value asymptotes. Quantitatively, the weak-to-strong measurement transition is characterized by a universal transition factor $e^{-\Gamma^2/2}$, where $\Gamma$ is a dimensionless parameter related to the system–apparatus coupling. This transition, which continuously connects weak measurements and strong measurements, may open new experimental possibilities to test quantum foundations and prompt us to re-examine and improve the measurement schemes of related quantum technologies.**

To date, quantum mechanics has succeeded in describing a variety of physical, chemical and even biological phenomena with unprecedented precision. However, fundamental challenges remain, for example, the quantum measurement problem is still considered to be an unsolved puzzle, persisting from the birth of quantum mechanics. The mathematical formalism of quantum measurement was set forth by von Neumann in 1932[1], by treating both the measured system and measuring apparatus as being quantum and coupling them through a simple interaction Hamiltonian. This process, also called 'pre-measurement', is unitary and is followed by a non-unitary macroscopic amplification that selects only a single outcome $a_n$, being an eigenvalue of the measured operator $\hat{A}$ with eigenstate $|n\rangle$. By repeating this measurement procedure over a large ensemble of similarly prepared systems with state $|i\rangle$, the expectation value of $\hat{A}$ can be represented as $\langle A \rangle_i \equiv \langle i|\hat{A}|i\rangle/\langle i|i\rangle = \sum_n |\langle i|n\rangle|^2 \langle A \rangle_n$, where $\langle A \rangle_n = \langle n|\hat{A}|n\rangle/\langle n|n\rangle$. Then, in 1988, von Neumann's measurement scheme, the so-called projective ('strong') measurement, was extended by Aharonov, Albert and Vaidman to the weak-coupling regime, corresponding to the situation where the measuring apparatus interacts weakly with the measured system[2]. A weak measurement can only obtain a small amount of information about the measured system. However, by increasing the number of measurement trials, we can still approach $\langle A \rangle_i$ with any desired precision.

In ref. [2], Aharonov, Albert and Vaidman also showed that, by introducing a post-selection procedure, a weak measurement enables one to record information regarding a pre- and post-selected ensemble using the concept of a 'weak value'. This procedure requires both pre-selection (that is, $|i\rangle$, the prepared initial state) and post-selection (that is, $|f\rangle$, the final state being strongly projected or filtered) of the measured system as shown in Fig. 1a.

The weak value is defined as $\langle A \rangle_w = \langle f|\hat{A}|i\rangle/\langle f|i\rangle$, where $\hat{A}$ is the measured observable. Note that the weak value is, in general, a complex number and can sometimes be 'superweak'[3] or 'anomalous'[4–8] lying well outside the spectrum of the measured operator. This has led to successful demonstrations of weak-value amplification techniques[9–11]. However, anomalous weak values demand the price of a small success rate of post-selection $p = |\langle f|i\rangle|^2$ due to the approximate orthogonality of the pre- and post-selected states. Alongside practical applications, weak values and weak measurements have also been linked to many fundamental topics and quantum paradoxes, in theory and experiment[12–18]. When the measured system and the measuring apparatus interact strongly, the post-selection will offer a measurement outcome characterized by the expectation-value asymptotics. Note that the weak value is quantitatively and conceptually different from the expectation value[19,20]. Consequently, the observation of these two values can provide a direct test of whether the performed measurement is weak or strong, especially in the context of experimental realization[19–24].

A natural question to ask is how does the measurement outcome transition from the weak value to the expectation value when the system–apparatus interaction is tuned from weak to strong? In this work, we experimentally demonstrate the weak-to-strong measurement transition using a single trapped $^{40}Ca^+$ ion to unify the weak-value and expectation-value predictions. The trapped-ion system has been widely developed as one of the most promising candidates for achieving large-scale quantum simulation and computation[25–27]. In this context, it offers us a well-designed measurement set-up with variable measurement strength. The ion's internal states and vibrational motion are identified as the measured system and measuring apparatus, respectively. Following von Neumann's measurement scheme[1,2], we introduce a generic system–apparatus coupling Hamiltonian $\hat{H}_I = \gamma_0 \hat{A} \hat{p}$ using a bichromatic light field, where the measured observable $\hat{A}$ and the momentum operator $\hat{p}$ belong to the system and the apparatus, respectively. By tuning the system–apparatus coupling time $t$, we can control the measurement strength $\gamma_0 t$ in our experiment. After the pre-selection step, the system–apparatus coupling, and the post-selection step, we finally generate a quantum superposition of two wavepackets of the apparatus (Fig. 1e). Its central-position shift relative to the initial state (the vibrational ground state) gives us the measurement outcome transitioning from the weak value to the expectation value when the measurement strength is tuned.

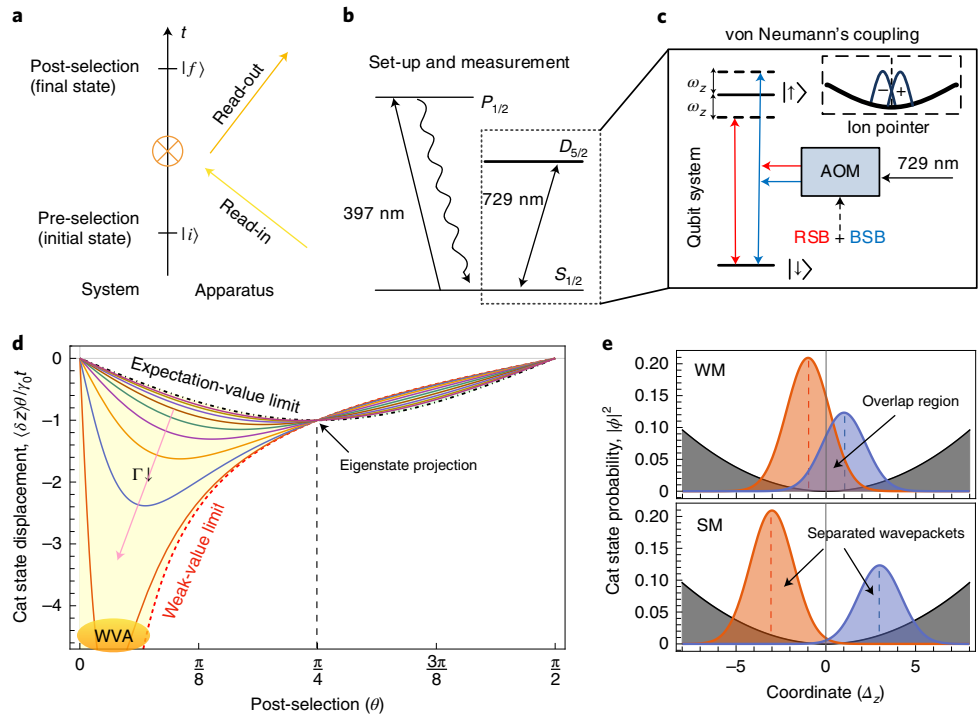

**Fig. 1 | The weak-to-strong quantum measurement set-up in a trapped-ion system and its weak-value to expectation-value transition. a**, The measurement procedure consists of pre-selection, system–apparatus coupling and post-selection steps. **b,c**, The single trapped $^{40}$Ca$^+$ ion can offer a well-designed measurement set-up with variable measurement strength with its internal electronic states and axial vibrational motion playing the roles of the measured system and measuring apparatus, respectively. The acousto-optical modulators (AOM) are used as frequency shifters and switches of the laser pulses with both the red- and blue-sideband (RSB and BSB) transitions. **d**, The relative spatial displacement of the ion's motional state $\langle \delta z \rangle_\theta / \gamma_0 t$ as a function of the post-selection angle $\theta$ with different $\Gamma = \gamma_0 t / \Delta_z$. The weak-value amplification (WVA) emerges in the nearly orthogonal regime $\theta \to 0$, and the weak-coupling limit $\Gamma \to 0$. **e**, Two spatially superposed wavepackets in the motional cat state overlap, with an interference contribution, in the weak-measurement (WM) regime (top), but are well separated in the strong-measurement (SM) regime (bottom). Gray shaded regions represent the potentials for quantizing the ion's motional state into harmonic oscillators (that is, phonon modes).

### Set-up and modelling

In our experimental demonstration (see Supplementary Section 1 for further details), the Zeeman sublevels $S_{1/2}(m_J = -1/2)$ and $D_{5/2}(m_J = -1/2)$ of a single trapped $^{40}$Ca$^+$ ion in a magnetic field of 5.3 G are chosen as system states $|\downarrow\rangle$ and $|\uparrow\rangle$, which compose a qubit with energy spacing $\hbar \omega_0$. The resonant transition between $|\downarrow\rangle$ and $|\uparrow\rangle$, that is, the so-called carrier transition, is realized using a narrow-linewidth laser at 729 nm. The ion's axial vibrational motion (along the $z$ direction), which is treated as a quantum harmonic oscillator with a frequency of $\omega_z = 2\pi \times 1.41$ MHz, is chosen as the measuring apparatus. Its ground-state wavepacket has the size $\Delta_z = \sqrt{\frac{\hbar}{2m\omega_z}} = 9.47$ nm, where $m$ is the ion's mass. Laser fields with frequencies $\omega_0 - \omega_z$, $\omega_0 + \omega_z$ can induce the so-called red-sideband and blue-sideband transitions, which are related to $|\uparrow\rangle\langle\downarrow|\hat{a} + \text{h.c.}$ and $|\uparrow\rangle\langle\downarrow|\hat{a}^\dagger + \text{h.c.}$ respectively, where $\hat{a}^\dagger$ and $\hat{a}$ are the creation and annihilation operators for the vibrational motion. The von Neumann coupling between the measured system and the measuring apparatus is realized by a bichromatic laser beam simultaneously resonant with both the red- and blue-sideband transitions[26,28], as depicted in Fig. 1b,c (see Methods and Supplementary Fig. A1 and Supplementary Section 2). The corresponding interaction Hamiltonian in the Lamb–Dicke approximation reads

$$\hat{H}_I = \gamma_0 \hat{\sigma}_x \hat{p}. \tag{1}$$

Here $\hat{\sigma}_x$ is the Pauli-$x$ operator of the qubit and $\hat{p}$ is the momentum operator of the oscillator. The coupling parameter is $\gamma_0 = \eta \Omega \Delta_z$ with the Lamb–Dicke parameter $\eta = 0.08$ and the Rabi frequency $\Omega = 2\pi \times 19$ kHz. A third rapidly decaying level $P_{1/2}$ (lifetime about 7.1 ns) is used for laser cooling and qubit state readout with a laser field at 397 nm as shown in Fig. 1b. Also, the wavepacket and the average spatial displacement of the axial motion after post-selection are measured indirectly by means of photon fluorescence detection (see Methods and Supplementary Section 4).

Using Doppler cooling, resolved sideband cooling and optical pumping[27,28], the internal electronic state of the ion is initialized in $|i\rangle = |\downarrow\rangle$ and the axial vibrational motion is prepared in the ground state $|\phi(z)\rangle = \left(\frac{1}{2\pi\Delta_z^2}\right)^{\frac{1}{4}} \exp\left(\frac{-z^2}{4\Delta_z^2}\right)$. After applying the system–apparatus interaction in equation (1) for a controllable time duration $t$, we post-selected the qubit system in the final state $|f\rangle = \cos\theta |\uparrow\rangle - \sin\theta |\downarrow\rangle$ by a projective measurement (see Methods and Supplementary Section 3). On the Bloch sphere of the qubit

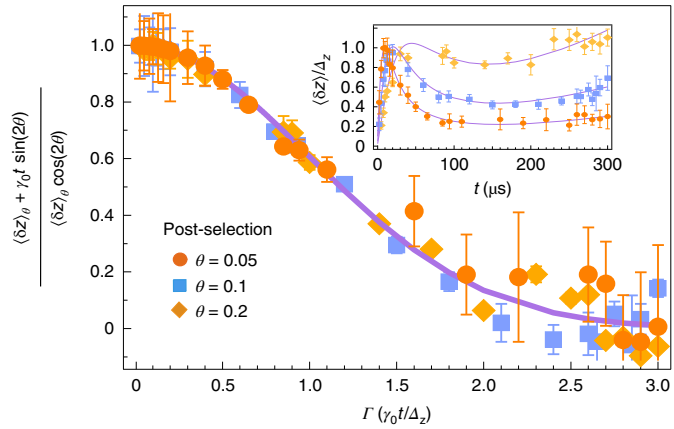

**Fig. 2 | The weak-to-strong measurement transition is exactly characterized by the overlapping extent of the motional cat state's superposed wavepackets** $\langle\phi(z+\gamma_0 t)|\phi(z-\gamma_0 t)\rangle = e^{-\Gamma^2/2}$. Based on the experimentally measurable pointer's spatial displacement $\langle\delta z\rangle_\theta$, this transition factor can be observed indirectly for different $\theta$ using $e^{-\Gamma^2/2} = (\langle\delta z\rangle_\theta + \gamma_0 t \sin(2\theta))/(\langle\delta z\rangle_\theta \cos 2\theta)$. The purple curve is the theoretical prediction. The points are experimental data. Inset: the ion's axial displacement $\langle\delta z\rangle_\theta/\Delta_z$. The agreement between prediction and data indicates that the weak-to-strong measurement transition has a universal transition factor regardless of the post-selection procedure. The error bars represent the fitting errors using the weighted fitting method.

system, the vector angle between the pre- and post-selected states is $\pi - 2\theta$. Considering the pre- and post-selection performed on the system, we can anticipate the measurement outcome of the Pauli operator $\hat{\sigma}_x$ in both the weak- and strong-coupling configurations. In the context of the weak-measurement regime, the weak value of the Pauli-$x$ observable is obtained as

$$\langle\sigma_x\rangle_W = \frac{\langle f|\hat{\sigma}_x|i\rangle}{\langle f|i\rangle} = -\cot\theta. \quad (2)$$

On the other hand, in the strong-measurement regime, the expectation value is given by

$$\langle\sigma_x\rangle_S = \frac{\langle f|\hat{\sigma}_x|f\rangle}{\langle f|f\rangle} = -\sin 2\theta. \quad (3)$$

Usually, the weak value in equation (2) is not equal to the expectation value in equation (3), except for the eigenstate projection (for example, at $\theta = \pi/4$). If we represent the qubit state as a vector on the Bloch sphere, the post-selected final state is orthogonal at angle $\theta = 0$ and parallel at $\theta = \pi/2$ to the initial state. Again, we claim based on our experimental demonstration that the essential difference between the weak- and strong-measurement schemes is completely captured by the observable's measurement outcome, being a weak value in equation (2) or an expectation value in equation (3).

To verify the above expectation and claim, the measurement process as depicted in Fig. 1a–c can be further described in an exact mathematical form

$$\underbrace{\langle f|}_{\text{post-selection}} \underbrace{\exp(-\frac{i}{\hbar}\int_0^t \hat{H}_I(t')dt')}_{\text{von Neumann coupling}} \underbrace{|i\rangle \otimes |\phi(z)\rangle}_{\text{pre-selection}} \quad (4)$$
$$= -\left(\frac{\sin\theta+\cos\theta}{2}\right)|\phi(z+\gamma_0 t)\rangle + \left(\frac{\cos\theta-\sin\theta}{2}\right)|\phi(z-\gamma_0 t)\rangle.$$

After normalization, the final pointer state of the ion's vibrational motion has a form of the cat state as

$$|\text{cat}_\theta\rangle = \frac{-\sin(\theta+\frac{\pi}{4})|\phi(z+\gamma_0 t)\rangle + \cos(\theta+\frac{\pi}{4})|\phi(z-\gamma_0 t)\rangle}{\sqrt{1-\cos(2\theta)\langle\phi(z+\gamma_0 t)|\phi(z-\gamma_0 t)\rangle}}. \quad (5)$$

By defining a dimensionless factor $\Gamma = \gamma_0 t/\Delta_z$, that is, the ratio between the interaction strength and the measuring pointer's width, the overlap between the two spatially separated Gaussian wavepackets of the cat state, which contributes to the quantum interference effect in this measurement procedure, can be quantified as

$$\langle\phi(z+\gamma_0 t)|\phi(z-\gamma_0 t)\rangle = e^{-\Gamma^2/2}. \quad (6)$$

Therefore, $\Gamma$ can be simply regarded as the 'interference factor' in our measurement setting, which plays an important role in the transition from weak to strong (projective) measurement.

The central-position shift of the motional cat state relative to the initial ground state, corresponding to the measurement outcome, can be obtained as

$$\langle\delta z\rangle_\theta = \frac{\langle\text{cat}_\theta|\hat{z}|\text{cat}_\theta\rangle}{\langle\text{cat}_\theta|\text{cat}_\theta\rangle} = -\frac{\gamma_0 t \sin 2\theta}{1-\cos(2\theta)e^{-\Gamma^2/2}}. \quad (7)$$

In the weak-coupling regime $\Gamma \ll 1$, we found $\langle\delta z\rangle_\theta|_{\Gamma\to 0} = -\gamma_0 t \cot\theta = \gamma_0 t\langle\sigma_x\rangle_W$, corresponding to the weak value of the measured Pauli operator (equation (2)). Whereas in the strong-coupling regime $\Gamma \gg 1$, we obtain $\langle\delta z\rangle_\theta|_{\Gamma\to\infty} = -\gamma_0 t \sin(2\theta) = \gamma_0 t\langle\sigma_x\rangle_S$, in accordance with the expectation value (equation (3)). The correspondence in both limits is not a coincidence, it reveals that the process of quantum measurement corresponds to creating entanglement between the measured system and the measuring apparatus via von Neumann coupling. The interference factor $\Gamma$ affects the outcomes of the measurement, and as a result, determines the weak value and expectation value in a unified way (equation (7)). At the selection angle $\theta = \pi/4$, we notice that $\langle\delta z\rangle_{\pi/4} = -\gamma_0 t$ and $\langle\sigma_x\rangle_W = \langle\sigma_x\rangle_S$, since the final apparatus state reduces to a single shifted wavepacket $|\phi(z+\gamma_0 t)\rangle$, and correspondingly the final qubit system state, that is $|-\rangle = (|\uparrow\rangle - |\downarrow\rangle)/\sqrt{2}$, is one eigenstate of the Pauli-$x$ operator.

## Weak value versus expectation value

Figure 1d shows the relative spatial displacement of the ion's motional state $\langle\delta z\rangle_\theta/\gamma_0 t$ as a function of the post-selection angle $\theta$ with different $\Gamma$. As indicated by the correspondence between $\langle\delta z\rangle_\theta/\gamma_0 t$ of the pointer and $\langle\sigma_x\rangle$ of the system, all the possible outcomes of the measured observable, lying in between the weak-value asymptotics (red dashed line) and the expectation-value asymptotics (black dot-dashed line), are presented in the shaded region of Fig. 1d. When the interference factor is decreased $\Gamma \to 0$, the relative spatial displacement shows 'blow-up' shift characteristics that indicate an anomalous weak-value amplification at the nearly orthogonal post-selection regime $\theta \to 0$. This anomalous phenomenon exhibited by the weak measurement has been applied to the detection of extremely weak signals, such as the spin Hall effect of light[9], and could provide practical advantages in the presence of detector saturation and technical noise[29,30].

Figure 1e explains the measurement results in both the limits of weak measurement and strong measurement based on the properties of the cat state. Note that the overlap of the two superposed wavepackets is essential for the survival of the weak-value amplification. Under the condition $\gamma_0 t > \Delta_z$, the overlap effect diminishes, and the two wavepackets are well separated, eventually resulting in the expectation value. Both the interaction strength $\gamma_0 t$ and the size (standard deviation) of the measuring pointer $\Delta_z$ are relevant to determine the weak-to-strong measurement transition. The ratio

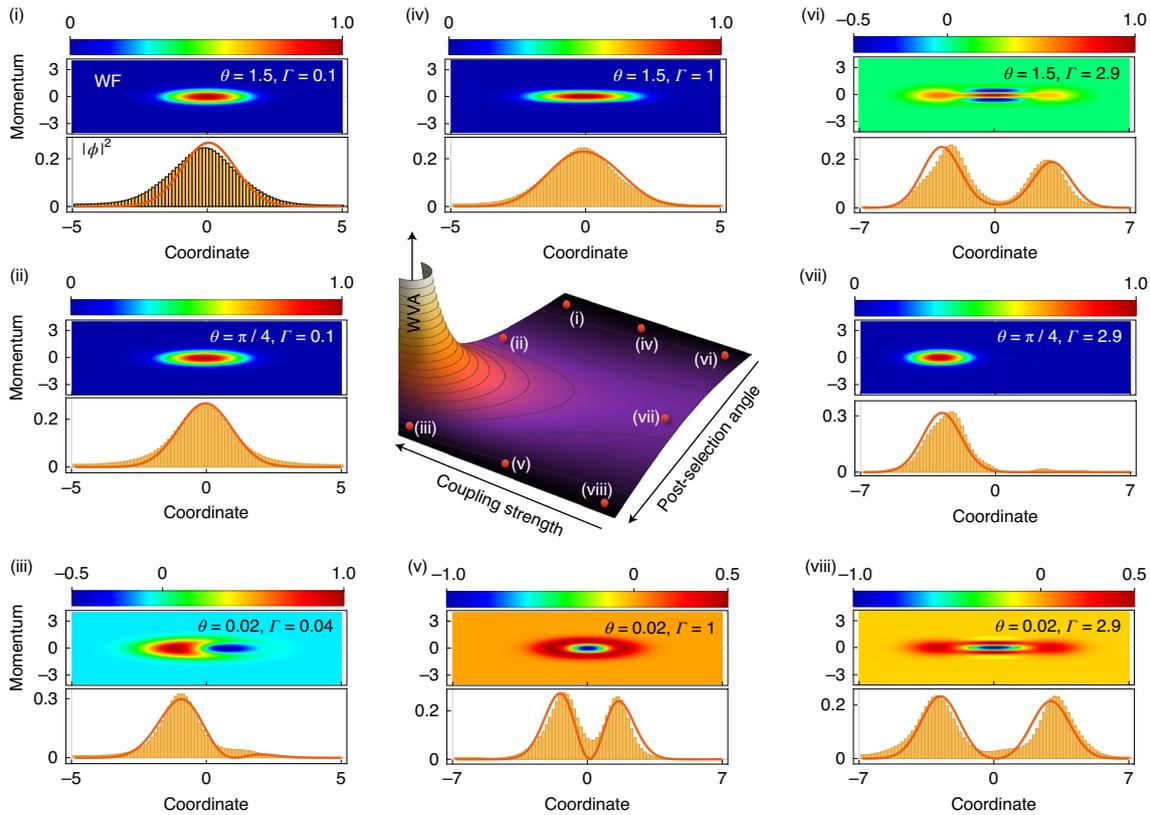

**Fig. 3 | The measurement regimes of $\langle \delta z \rangle_\theta /(\gamma_0 t)$ in the full parameter space $(\Gamma, \theta)$.** The probability density distributions of the typical cat states are reconstructed in the weak-measurement regime with coupling strength $\Gamma = 0.1, 0.04$ ((i)–(iii)), the intermediate-measurement regime with $\Gamma = 1$ ((iv),(v)) and the strong-measurement regime with $\Gamma = 2.9$ ((vi)–(viii)). The post-selection angles $\theta = 0.02, \pi/4, 1.5$ correspond to nearly orthogonal, eigenstate projection and nearly parallel between the pre- and post-selected states, respectively. The Wigner functions (WF) are plotted based on the theoretical results (see Methods) using $\Delta_z$ and $\hbar/2\Delta_z$ as the units of coordinate and momentum. For the probability density distributions, the difference between the theoretical predictions (red curves) and the experimental data (yellow histograms) originates from the spin-state detection error and the instability of experimental parameters during the long timescale of data acquisition.

between these two parameters can characterize the measurement regime adequately.

### The weak-to-strong measurement transition

The weak-to-strong transition of measurement is exactly characterized by the transition factor $e^{-\Gamma^2/2}$, which quantifies the overlap of the two wavepackets and shows a general law for weak-to-strong transition in different coupling regimes that were preliminarily investigated in previous works[31–33]. In experiment, this transition property is difficult to observe directly. However, based on the pointer's experimentally measurable spatial displacement $\langle \delta z \rangle_\theta$ under different post-selection angles $\theta$ and ratios $\Gamma$, we can infer information on the transition factor by rewriting equation (7) as $e^{-\Gamma^2/2} = (\langle \delta z \rangle_\theta + \gamma_0 t \sin(2\theta))/(\langle \delta z \rangle_\theta \cos 2\theta)$. Figure 2 compares the theoretical prediction (purple curve) and the experimental data (points) as a function of $\Gamma$ ranging from the weak-coupling regime ($\Gamma \sim 0.02$) to the strong-coupling regime ($\Gamma \sim 3.0$). The nearly perfect agreement between the theoretical prediction and the experimental data indicates that the weak-to-strong transition is universal for the Gaussian apparatus, regardless of the specific pre- and post-selections of the measured qubit system[19]. It should be emphasized that for other types of measuring pointer, for example, Lorentzian pointer and exponential pointer[34–37], the transition factors have different mathematical forms.

Because of the critical role of $\Gamma = \gamma_0 t/\Delta_z$ in determining the measurement regime, the measurement transition can be implemented experimentally in three ways, each corresponding to tuning one of the three parameters $\gamma_0$, $t$ and $\Delta_z$. The parameters $\gamma_0$ and $t$ determine the coupling strength and duration between the qubit and the ion's axial vibrational motion, respectively. More interestingly, since $\Gamma$ is inversely proportional to the wavepacket's width, $\Delta_z$ determines the sensitivity of the measuring apparatus. The decrease in $\Delta_z$, implying the high sensitivity of the measuring apparatus and the rapid loss of the wavepackets' overlap, leads to easy access to the strong-measurement regime. In a single-trapped-ion system, tuning the magnitude of $\Delta_z$ can be realized by applying the squeezing technique to the ion's axial motion, as discussed in refs. [25,38]. Nevertheless, the most straightforward approach to implement the weak-to-strong transition is to tune the coupling duration $t$ as we did in this work.

### State reconstruction of measurement regimes

In Fig. 3, we demonstrate the measurement regimes in the full parameter space $(\Gamma, \theta)$. The typical superposition states in the weak (also called kitten state), intermediate and strong (cat state) measurement regimes are reconstructed.

Figure 3(i)–(iii) present the probability density distributions and Wigner functions of the pointer states in the weak-measurement regime with $\Gamma = 0.1, 0.04$ and post-selection angles $\theta = 0.02, \pi/4, 1.5$. The three angles correspond to nearly orthogonal, eigenstate projection and nearly parallel pre- and post-selected states, respectively. Note that the remarkable weak-value amplification

appears in the nearly orthogonal post-selection case, and the pointer shows a distinct shift in the position space. As shown in Fig. 3(iii), we have amplified a tiny spatial displacement of ~4 Å ($\Gamma = 0.04$) of the single trapped ion to ~10 nm by post-selecting the appropriate internal electronic qubit state in the nearly orthogonal case $\theta = 0.02$. That is, signal amplification by a factor of 25 is reported (also see ref. [28]). Note that the trade-off for achieving such a large amplification is the small success probability $p \approx \sin^2\theta \approx 4 \times 10^{-4}$.

For $\Gamma = 1$, the transition factor is $e^{-\Gamma^2/2} \sim 0.6$, and there is still a distinct overlap between the two superposed wavepackets. Figure 3(iv),(v) give the probability density distributions and Wigner functions of the pointer states in this intermediate-measurement regime with post-selection angles $\theta = 0.02, 1.5$. The rise (Fig. 3(iv)) and dip (Fig. 3(v)) of the probability density at the zero coordinate come from the constructive and destructive interferences of the overlapped wavepackets, respectively. By choosing $\Gamma = 2.9$, the transition factor $e^{-\Gamma^2/2} \sim 0.01$ and the two wavepackets have negligible overlap, and then we enter the strong-coupling regime. Figure 3(vi)–(viii) reconstruct the cat states in this regime with $\theta = 0.02, \pi/4, 1.5$. For the nearly parallel (Fig. 3(vi)) and the nearly orthogonal (Fig. 3(viii)) post-selection cases, the corresponding constructive and destructive interferences between the two well-separated wavepackets will give similar probability density distributions of the cat states. At $\theta = \pi/4$, the eigenstate projection occurs and only a single wavepacket shifts to the left.

## Conclusion

In summary, the weak-to-strong measurement transition predicts the continuous relation between the weak-value asymptotics and expectation-value asymptotics. Full control of this transition was demonstrated by using a single trapped $^{40}$Ca$^+$ ion. The quantum measurement process is investigated in detail by taking the ground state of the ion's axial motion as the measuring apparatus. The transition factor $\exp(-\Gamma^2/2)$ is universal for Gaussian-type apparatuses and was also found in the classical-to-quantum transition in the context of light–matter interactions[39]. Prospectively, the measurement transition from weak values to expectation values offers a promising avenue to re-examine fundamental topics and core issues related to quantum measurements (perhaps even the measurement problem itself), and may also highlight the versatility of trapped-ion systems for future applications.

## Online content

Any methods, additional references, Nature Research reporting summaries, source data, extended data, supplementary information, acknowledgements, peer review information; details of author contributions and competing interests; and statements of data and code availability are available at https://doi.org/10.1038/s41567-020-0973-y.

## Methods

**Interaction between the measured system and the measuring pointer.** In our experiment, the two energy states, that is, $S_{1/2}(m_J = -1/2)$ and $D_{5/2}(m_J = -1/2)$, and the axial motional states of a single $^{40}\text{Ca}^+$ ion are chosen as the system and pointer degrees of freedom, respectively. Three basic quantum operations are used in our measurement procedure:

$$\begin{aligned}\hat{H}_{\text{car}} &= \frac{\hbar\Omega}{2}(\hat{\sigma}^+ e^{i\phi_{\text{car}}} + \hat{\sigma}^- e^{-i\phi_{\text{car}}}), \\ \hat{H}_{\text{red}} &= \frac{i\hbar\eta\Omega}{2}(\hat{a}\hat{\sigma}^+ e^{i\phi_{\text{red}}} - \hat{a}^\dagger\hat{\sigma}^- e^{-i\phi_{\text{red}}}), \\ \hat{H}_{\text{blue}} &= \frac{i\hbar\eta\Omega}{2}(\hat{a}^\dagger\hat{\sigma}^+ e^{i\phi_{\text{blue}}} - \hat{a}\hat{\sigma}^- e^{-i\phi_{\text{blue}}}),\end{aligned} \quad (8)$$

where $\hat{H}_{\text{car}}$, $\hat{H}_{\text{red}}$ and $\hat{H}_{\text{blue}}$ are the Hamiltonians of the carrier, the red sideband and the blue sideband transitions, $\phi_{\text{car}}$, $\phi_{\text{red}}$ and $\phi_{\text{blue}}$ are the corresponding laser phases, $\eta$ is the Lamb–Dicke parameter, $\Omega$ is the Rabi frequency, and $\hat{a}^\dagger$ and $\hat{a}$ are the creation and annihilation operators for the motional degree of freedom respectively. A bichromatic laser field resonant with the blue and red sidebands of the ion gives a combined operation $\hat{H}_{\text{bic}} = \hat{H}_{\text{red}} + \hat{H}_{\text{blue}}$. The interaction Hamiltonian can be recast into a new form

$$\hat{H}_{\text{bic}} = \frac{\hbar\eta\Omega}{2}[\hat{\sigma}_x \sin\phi_+ + \hat{\sigma}_y \cos\phi_+] \otimes [-(\hat{a}^\dagger + \hat{a})\cos\phi_- + i(\hat{a}^\dagger - \hat{a})\sin\phi_-], \quad (9)$$

where $\phi_+ = (\phi_{\text{red}} + \phi_{\text{blue}})/2$ and $\phi_- = (\phi_{\text{red}} - \phi_{\text{blue}})/2$ indicate the sum and difference of the red-sideband phase and the blue-sideband phase. It can be seen that the system operator is determined by $\phi_+$ and the pointer operator is selected by $\phi_-$.

By setting $\phi_+ = \pi/2$, $\phi_- = \pi/2$, the interaction Hamiltonian is simplified to $\hat{H}_{\text{bic}} = \eta\Omega\Delta_z\hat{\sigma}_x\hat{p}$, where $\hat{p} = \frac{i\hbar(\hat{a}^\dagger - \hat{a})}{2\Delta_z}$, which is the von Neumann coupling we have used throughout this work (equation (1)). When we choose $\phi_+ = \pi/2$, $\phi_- = \pi$, equation (9) reduces to

$$\hat{H}_{\text{bic}} = \frac{\hbar\eta\Omega}{2}\hat{\sigma}_x(\hat{a} + \hat{a}^\dagger) = \frac{\hbar\eta\Omega/\Delta_z}{2}\hat{\sigma}_x\hat{z}. \quad (10)$$

The corresponding unitary evolution operator is

$$\hat{U}_z = \exp(-ik\hat{z}\hat{\sigma}_x/2), \quad (11)$$

where $k = \eta\Omega t/\Delta_z$. The evolution operator (equation (11)) was used in the reconstruction of the motional wavepacket and the measurement of $\langle\hat{z}\rangle$.

**Pre- and post-selection.** In the experiment, the pre-selected state $|\downarrow\rangle \otimes |\phi(z)\rangle$ was created by using the optical pumping and resolved sideband cooling techniques. The post-selected internal state can only be the dark state $|\uparrow\rangle$ with the motional state not destroyed, where no fluorescence is detected when irradiating with the 397 nm laser field. However, an arbitrary internal state $|f\rangle = \cos\theta|\uparrow\rangle - \sin\theta|\downarrow\rangle$ can effectively be post-selected by using a unitary rotation $R_y(2\theta)$ before post-selecting $|\uparrow\rangle$, where $R_y(2\theta)$ indicates the unitary rotation of a single qubit with angle $2\theta$ around the $y$ axis of the Bloch sphere. $R_y(2\theta)$ can be implemented with the carrier transition.

**Reconstruction of the motional wavepackets and measurement of $\langle\hat{z}\rangle$.** To obtain the information about the ion's final pointer state, the motional wavepacket's probability distribution should be measured. However, the only observable that can be measured directly for the trapped ion is $\hat{\sigma}_z$. With this in mind, we applied a unitary operation $\hat{U}_z = \exp(-ik\hat{z}\hat{\sigma}_x/2)$ (equation (11)) before measuring $\hat{\sigma}_z$, the effective measured observable will be

$$\hat{O}(k) = \hat{U}_z^\dagger \hat{\sigma}_z \hat{U}_z = \cos(k\hat{z})\hat{\sigma}_z + \sin(k\hat{z})\hat{\sigma}_y, \quad (12)$$

where $k = \eta\Omega t/\Delta_z$. It can be seen from this formula that $\langle\cos(k\hat{z})\rangle$ and $\langle\sin(k\hat{z})\rangle$ can be obtained by first preparing the internal state of the ion in the eigenstate of $\hat{\sigma}_z$ and $\hat{\sigma}_y$, respectively, and then measuring $\langle\hat{O}(k)\rangle$. The probability distribution for $\hat{z}$ can be extracted by using the Fourier transform of the function $g(k) = \langle\cos(k\hat{z})\rangle + i\langle\sin(k\hat{z})\rangle$, which is

$$G(z) = \int_{-\infty}^{\infty} g(k)e^{-ikz}\mathrm{d}k = \langle 2\pi\delta(\hat{z} - z)\rangle = 2\pi|\varphi(z)|^2, \quad (13)$$

where $|\varphi(z)|^2$ is the probability distribution of the measured wavepacket. However, there are some problems with using this method. Instead, we employed the constrained least-squares optimization method to obtain the probability distribution of the wavepacket (Supplementary Fig. A2 and Supplementary Section 4).

From equation (12) we have $\frac{\mathrm{d}}{\mathrm{d}k}\langle\hat{O}(k)\rangle|_{t=0} = \langle\hat{z}\hat{\sigma}_y\rangle$. By setting the internal state of the ion to the eigenstate of $\hat{\sigma}_y$ and probing $\langle\hat{O}(k)\rangle$ for different $k$, the expectation value of $\langle\hat{z}\rangle$ can be extracted via the slope of the data, see Supplementary Fig. A3 as an example.

**Wigner function distribution of the cat state.** In Fig. 3, we show the typical cat states of the vibrational motion in a phase-space representation with their Wigner functions. For a wavepacket $|\varphi(z)\rangle$, its Wigner function is defined as $W(z,p) = \frac{1}{2\pi\hbar}\int\varphi(z+z'/2)\varphi^*(z-z'/2)e^{-ipz'/\hbar}\mathrm{d}z'$. By substituting equation (5), one can obtain the Wigner function of the motional cat state

$$\begin{aligned}W(z,p) = (1 - \cos(2\theta)e^{-\Gamma^2/2})^{-1}\{&\sin^2(\theta + \pi/4)W^{(0)}(z + \gamma_0 t, p) \\ +\cos^2(\theta + \pi/4)W^{(0)}(z - \gamma_0 t, p) &- \cos(2\theta)\cos(2\gamma_0 pt)W^{(0)}(z,p)\},\end{aligned} \quad (14)$$

where the Wigner function $W^{(0)}(z,p) = \frac{1}{\pi\hbar}e^{-z^2/2\Delta_z^2 - 2\Delta_z^2 p^2/\hbar^2}$ describes the initial ground state of the ion's axial motion in $(z, p)$ phase space. And the spatial probability density distribution of the cat state can be obtained as $|\varphi(z)|^2 = \int W(z,p)\mathrm{d}p$.

## Data availability

Data that support the plots within this paper and other findings of this study are available from the corresponding authors upon reasonable request. Source data are provided with this paper.

## Acknowledgements

This work was supported in part by DIP (German–Israeli Project Cooperation) and by the I-CORE Israel Center of Research Excellence programme of the ISF and by the Crown Photonics Center, and was also supported by the National Basic Research Program of China under grant no. 2016YFA0301903 and the National Natural Science Foundation of China under grant nos. 61632021 and 11574398. E.C. acknowledges support from the Israeli Innovation Authority under project no. 70002 and from the Quantum Science and Technology Program of the Israeli Council of Higher Education.

## Author contributions

Y.P. and E.C. proposed the concept and the modelling. Y.P., J.Z., C.W. and P.C. contributed to the theoretical analysis, design and setting up of the experiments. J.Z. and C.W. performed the atom experiment, and analysed the data together with Y.P. and E.C. All authors contributed to the writing and revision of the manuscript.

## Competing interests

The authors declare no competing interests.

## Additional information

**Supplementary information** is available for this paper at https://doi.org/10.1038/s41567-020-0973-y.

**Correspondence and requests for materials** should be addressed to Y.P. or C.-w.W.

**Reprints and permissions information** is available at www.nature.com/reprints.